\documentclass{aa}
\usepackage{graphicx}
\usepackage{txfonts}
\begin{document}
\newcommand{\SCARGLE}{{\sffamily\scshape scargle\,}}
\newcommand{\MTRAP}{{\sffamily\scshape mtrap\,}}
\newcommand{\MOST}{{\sffamily\scshape MOST\,}}
\newcommand{\COROT}{{\sffamily\scshape CoRoT}}

\newcommand{\MgII}{\ion{Mg}{ii}}
\newcommand{\HeI}{\ion{He}{i}}
\newcommand{\Halpha}{H$\alpha$}
\newcommand{\Hbeta}{H$\beta$}
\newcommand{\Hgamma}{H$\gamma$}

\def\fu{$\nu_1$}
\def\fd{$\nu_2$}
\def\ft{$\nu_3$}
\def\cd{d$^{-1}$}
\def\cds{d$^{-1}$\,}
\def\kms{km~s$^{-1}$}

   \title{Multiperiodicity in the newly discovered mid-late Be star V2104\,Cygni\thanks{Based on observations obtained at the Observatorio Astron\'omico Nacional San Pedro M\'artir (Mexico), Observatorio de Sierra Nevada (Spain) and Observatoire de Haute Provence (France), and on observations made with the Nordic Optical Telescope, operated on the island of La Palma jointly by Denmark, Finland, Iceland, Norway, and Sweden, in the Spanish Observatorio del Roque de los Muchachos of the Instituto de Astrof\'{\i}sica de Canarias.}}

   \author{K.~Uytterhoeven \inst{1}
          \and 
           E.~Poretti \inst{1} 
	   \and
	   E.~Rodr\'{\i}guez \inst{2}
	   \and
           P.~De Cat \inst{3}
           \and
           P.~Mathias \inst{4}
           \and 
           J.H.~Telting \inst{5}
           \and
	   V.~Costa \inst{2}
           \and
           A.~Miglio \inst{6}
          }

   \offprints{K. Uytterhoeven}

   \institute{INAF-Osservatorio Astronomico di Brera, 
              Via E. Bianchi 46, I-23807 Merate\\
              \email{(katrien.uytterhoeven,ennio.poretti)@brera.inaf.it}
         \and
	 Instituto de Astrof\'{\i}sica de Andaluc\'{\i}a (CSIC), Apartado 3004, E-18080 Granada, Spain \\
         \email{(eloy,victor)@iaa.es}
         \and
	     Royal Observatory of Belgium, Ringlaan 3, B-1180 Brussel, Belgium\\
\email{peter@oma.be}
\and
Observatoire de la C\^ote d'Azur, GEMINI, CNRS, Universit\'e Nice
Sophia-Antipolis, BP 4229, F-06304 Nice Cedex 4, France\\
\email{Philippe.Mathias@oca.eu}
         \and
Nordic Optical Telescope, Apartado 474, E-38700 Santa Cruz de La Palma, Spain\\
\email{jht@not.iac.es}
\and
Institut d'Astrophysique et de G\'eophysique de l'Universit\'e de Li\`ege, All\'ee du 6 Ao\^ut 17, B-4000 Li\`ege, Belgium \\
\email{a.miglio@ulg.ac.be}
             }

   \date{Received ; accepted}

  \abstract {} {We obtained the first long, homogenous time-series of
  V2104\,Cyg, consisting of 679 datapoints, with the {{\sl
  uvby}}$\beta$ photometers of Sierra Nevada and San Pedro M\'artir
  Observatories with the aim to detect and subsequently interpret the
  intrinsic frequencies of this previously unstudied variable star,
  which turned out to be a Be star. We try to figure out its place
  among the variable B stars on the upper Main Sequence. In order to
  obtain additional information on physical parameters we collected a
  few spectra with the {{\sl Elodie}} and {{\sl FIES}}
  instruments.}{We searched for frequencies in the {{\sl uvby}}
  passbands using 2 different frequency analysis methods and used the
  S/N$>$4 criterion to select the significant periodicities.  We
  obtained an estimate of the physical parameters of the underlying B
  star of spectral type between B5 and B7, by correcting for the
  presence of a circumstellar disk, using a formalism based on the
  strenght of the \Halpha\, line emission.}{We detected 3 independent
  frequencies with amplitudes below 0.01\,mag, \fu$ = 4.7126$ \cds,
  \fd$ = 2.2342$ \cds and \ft$ = 4.671$ \cds, and discovered that
  V2104\,Cyg is a Be star.  The fast rotation (v$\sin
  i$=290$\pm$10 \kms, and 27\degr$<i<$ 45\degr) hampered the
  investigation of the associated pulsational parameters $\ell$. Nevertheless,
  the
  most plausible explanation for the observed variability of this
  mid-late type Be star is a non-radial pulsation model.}{}

   \keywords{Stars: oscillations -- Stars: emission-line, Be -- Stars: individual: V2104\,Cyg}

   \maketitle
%
%________________________________________________________________

\section{Introduction}
Several pulsators and variable stars of spectral type B occupy the
region of the upper Main Sequence. Among them are the $\beta$ Cephei
stars, the Slowly Pulsating B stars (SPBs) and the Be stars.

The former two classes are well-established pulsators that show
multiperiodic line-profile and light variations and whose pulsations
are driven by the $\kappa$-mechanism acting on the iron
bump. Most $\beta$ Cephei stars (spectral type B0.5--B3) pulsate in
low-order $p$- and $g$-modes with typical pulsation periods between 
0.3 and 0.8 days. The largest fraction of this pulsational class are
moderate rotators (v$\sin i < 150$ \kms). The SPBs are less massive B stars (spectral type
B2--B9) and slow rotators that pulsate in high radial order 
$g$-mode of low spherical harmonic degree $\ell$ with typical pulsation
periods of 0.5 to 5 days. We refer to Stankov \& Handler (2005) and De
Cat (2002) for overviews of the properties of the $\beta$ Cephei stars and SPBs, respectively.

The Be stars are a more complex and less well understood class of
near Main Sequence rapidly rotating B stars that produce a disk in their
equatorial plane and hence show Balmer line emission.  Be stars
exhibit variations on different time-scales (hours--years), with a
broad range of amplitudes.  Explanations for the variability have been
sought among the dynamics of the circumstellar disk, stellar wind,
rotational modulation, magnetic field and/or pulsations (e.g. Porter
\& Rivinius 2003). The highest fraction of Be stars have been found
among early-B type stars (B1--B2), while the observed cases in the
spectral type interval of B7--A2 are less frequent (Zorec 2000).
However, recently a number of late-type Be stars were discovered during the ground-based preparatory work of the
\COROT\, satellite (Neiner et al. 2005). The emission line B
stars occupy the instability strips of $\beta$ Cephei and SPB
pulsators and an obvious question arises whether or not the observed
short-period Be variations resemble the properties of these
well-established pulsators.

The short-periods observed in Be stars typically range from 0.2 to 5
days. Ground-based \emph{light curves} mainly reveal
\emph{monoperiodic} signals, with periods close to the rotational
period (the so-called $\lambda$ Eri variables, Balona 1990), and
amplitudes of a few tens of mmag.  These targets also show
line-profile variations (LPV) of low spherical harmonic degree, similar
to the LPV observed in $\beta$ Cephei variables, although the latter
have shorter periods. The variability has been attributed to
rotational modulation by Balona (1990, 1995) or non-radial pulsations
(NRP) (Baade 1982). Spectroscopically, nearly all Be stars of early-B
type show LPV (e.g.\, Rivinius et al. 2003). Contrary to ground-based
photometric data, the LPV of several early-type Be stars reveal multiperiodic
signals.  The observed variability is interpreted in terms of NRP
$g$-modes, typically with $\ell=m=+2$. Given the $g$-mode nature and
the similar length of the pulsation periods, these Be stars seem to
resemble the pulsational behaviour of SPBs. Until now, no clear cases
of LPV in late-Be type stars (later than B6) have been found (Baade
1989; Rivinius et al. 2003).  Recently, proof of the presence of
low-amplitude additional periods and variability of late-type Be stars
has been given by the \MOST satellite as \emph{multiperiodic} signals
have been detected in the light of the Be stars HD163868 (B5Ve, Walker
et al. 2005) and $\beta$ CMi (B8Ve, Saio et al. 2007). This result is
very encouraging and gives high expectations for the outcome of the Be
programme stars of the space mission \COROT, which has been launched
on December 27th 2006.

V2104\,Cyg (HD190397, HIP98611, $\alpha_{2000}$=$20^{\rm h} 01^{\rm m}
45.\!\!^{\rm s}54$, $\delta_{2000}$=$+57\degr 39\arcmin 06.\!\!''51$,
V=7.65\,mag) is listed in the HIPPARCOS catalogue (Perryman et al.\
1997) as an unsolved variable of spectral type A0, probably a
pulsating star, located at a distance of approximately $617 \pm 263$
pc. According to Grenier et al. (1999), the spectrum of V2104\,Cyg
resembles a B8 Main Sequence star. Such a late-type B star or
early-type A star would be a challenging object for many variable
classes.  In order to clarify this vague characterization we obtained
a photometric time-series and a few spectra. From the spectra it soon
became evident that V2104\,Cyg\, shows Balmer line emission (see
Sect.~\ref{spectra}). The study of V2104\,Cyg is particularly
interesting as (multiperiodic) variability is scarcely detected in
late-type Be stars (see above). We present an analysis of the variable
character of this newly discovered Be star and try to figure out its
place among the variable B stars on the upper Main Sequence.

%__________________________________________________________________

\section{Photometric observations \label{sect2}}

\begin{figure*}
\centering
\resizebox{0.98\linewidth}{!}{\rotatebox{90}{\includegraphics{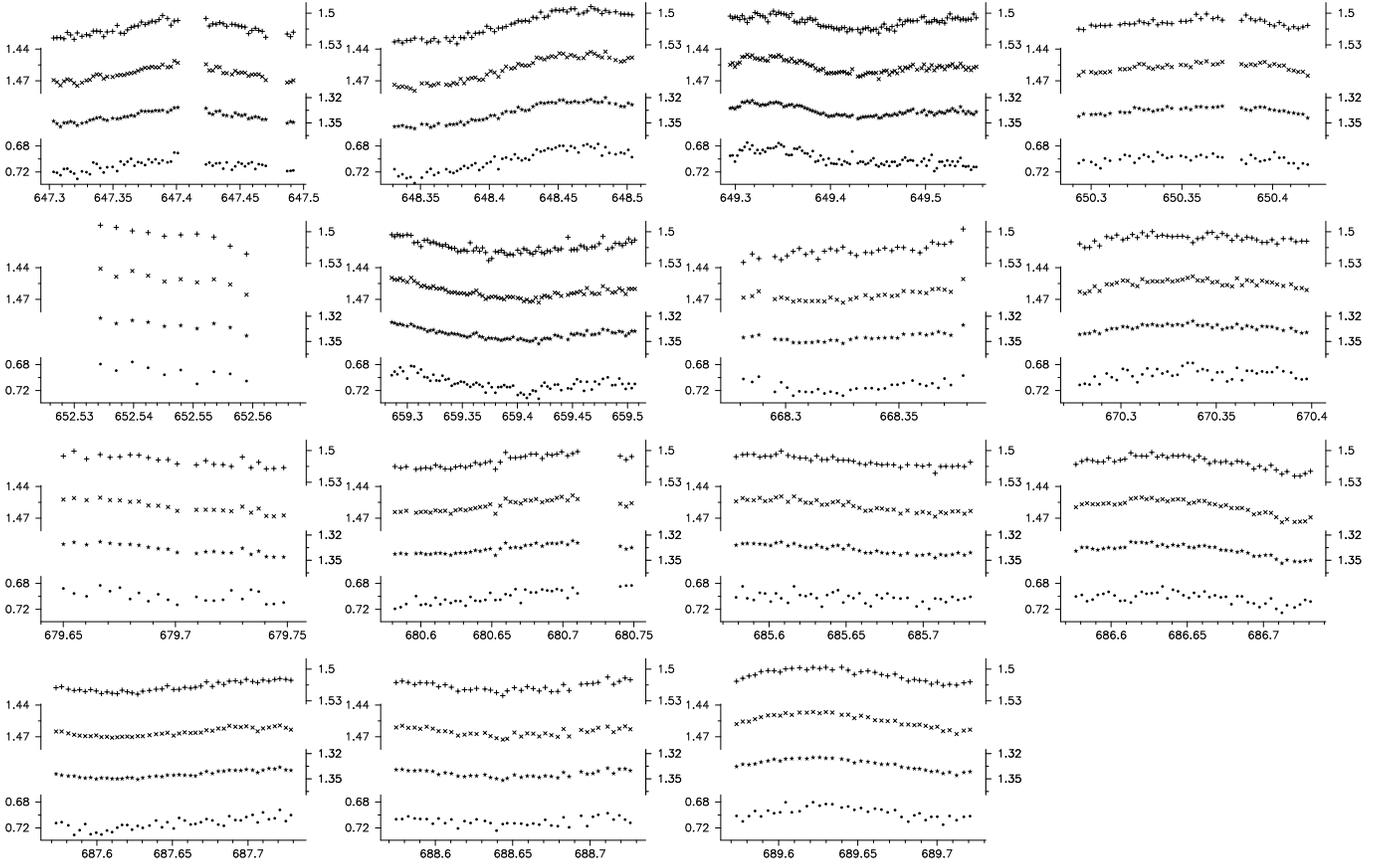}}}
\caption{Light curves of V2104\,Cyg obtained at SNO and SPMO  in
  October--November 2005. For each night (from bottom to top) the {\sl u} (dots),
  {\sl v} (stars), {\sl b} (crosses) and {\sl y} (+) magnitudes are plotted in time
  (HJD-2453000). Note the difference in scale between the {\sl u} and {\sl vby}
  data: tickmarks in the {\sl u} filter and {\sl vby} filters correspond to
  0.04\,mag and 0.03\,mag respectively.}
\label{lightcurves}
\end{figure*}

V2104\,Cyg was observed in the framework of a double-site campaign in
the autumn of 2005. Two twin Danish six-channel {\sl uvby}$\beta$
photometers were used at Sierra Nevada (SNO), Spain, and San Pedro
M\'{a}rtir (SPMO), Mexico, observatories, attached to the 90 cm and
1.5 m telescopes, respectively.  Both photometers are equipped for
simultaneous measurements in {\sl uvby} or the narrow and wide \Hbeta\,
channels (Nielsen 1983). Nevertheless, most of the data were collected
in the four {\sl uvby} filters in order to investigate the variability
behaviour of this star. Only a few points were obtained in \Hbeta\, at
SNO for calibration purposes of the photometric indices.  

HR~7634 ($V$=6.16\,mag, A4Vn) and HR~7692 ($V$=6.19\,mag, F4V) were used
as comparison stars in both sites. A total of 414 {\sl uvby}
datapoints of V2104\,Cyg were collected on 8 nights at SNO, from
October 3 to 26, and 265 {\sl uvby} datapoints on 7 nights at SPMO,
from November 4 to 14. In total, about 54 hours of useful data were
acquired. The light curves are given in Fig.~\ref{lightcurves}.

The quality of the nights was excellent at SPMO. Bouguer's lines,
i.e. the least-squares fit to the observations of a same star versus
airmass for each night, yielded correlation coefficients in the
interval 0.994--1.000, whereby 23 times out of 28 in the 0.998--1.000
interval. Mean extinction coefficients were 0.472\,mag in {\sl u}
light, 0.267\,mag in {\sl v} light, 0.165\,mag in {\sl b} light and
0.116\,mag in {\sl y} light. Similar mean coefficients, with standard deviation given between  brackets, were obtained
at SNO: 0.484\,mag (0.013\,mag), 0.278\,mag
(0.009\,mag), 0.171\,mag (0.007\,mag) and 0.121\,mag
(0.010\,mag) for filters {\sl u, v, b} and {\sl y},
respectively, with the airmass ranging from about 1.05 to 1.5 each
observing night.

An analysis of the magnitude differences between the two comparison stars was
carried out to get an additional feeling of the quality of the dataset. The
overall standard deviations of the 268 magnitude differences obtained at SPMO
are: 6.7\,mmag in {\sl u} light, 2.0\,mmag in {\sl v} light, 1.9\,mmag in {\sl b} light
and 1.8\,mmag in {\sl y} light.  The high scatter in the {\sl u} filter is due to some
leakage problems in the photomultiplier which affected the measurements in a random
way. The standard deviations of each night lie between 4.6--7.5, 1.4--2.3,
1.4--2.2 and 1.2--2.1\,mmag in {\sl u, v, b} and {\sl y} filters, respectively.

In case of the SNO data, we measured the two comparison stars 427
times and the magnitude differences yielded standard deviations of
6.3, 2.3, 2.3 and 2.9\,mmag in {\sl u, v, b} and {\sl y} filters,
respectively. Excluding the night JD 24453652 (where values are
1.5--2.0 times greater), the standard deviations of each night lie
between 5.5--6.7, 1.7--2.6, 1.4--2.3 and 2.1--2.7\,mmag in {\sl u, v,
b} and {\sl y} filters, respectively.

Since small differences are present in the mean values measured at SNO
an SPMO, the magnitude differences between the comparison stars have
been aligned. The combined timeseries have standard deviations of 6.4,
2.2, 2.1 and 2.3\,mmag in {\sl uvby}, respectively.  Therefore, our
photometric timeseries are characterized by a very good precision,
close to the best value we can obtain from ground.  The corresponding
periodogram is shown in the bottom panel of Fig.~\ref{scaperiodogram}.

\section{Frequency analysis}

The frequency search was carried out using Lomb-\SCARGLE
(Scargle 1982) as well as the least-squares power spectrum method
(Van\'{\i}\v{c}ek 1971). The latter method allows to detect the
constituents of the light curve one by one, whereby only the values of
the detected frequencies are introduced as known constituents in each
new search. Such a procedure differs from the \SCARGLE method
as it does not require any data prewhitening as amplitudes and
 phases of the known constituents are recalculated for each new trial
 frequency, whereby always the exact amount of signal for any detected
 term is subtracted.

\begin{figure*}
   \centering
\resizebox{0.95\linewidth}{!}{\rotatebox{-90}{\includegraphics{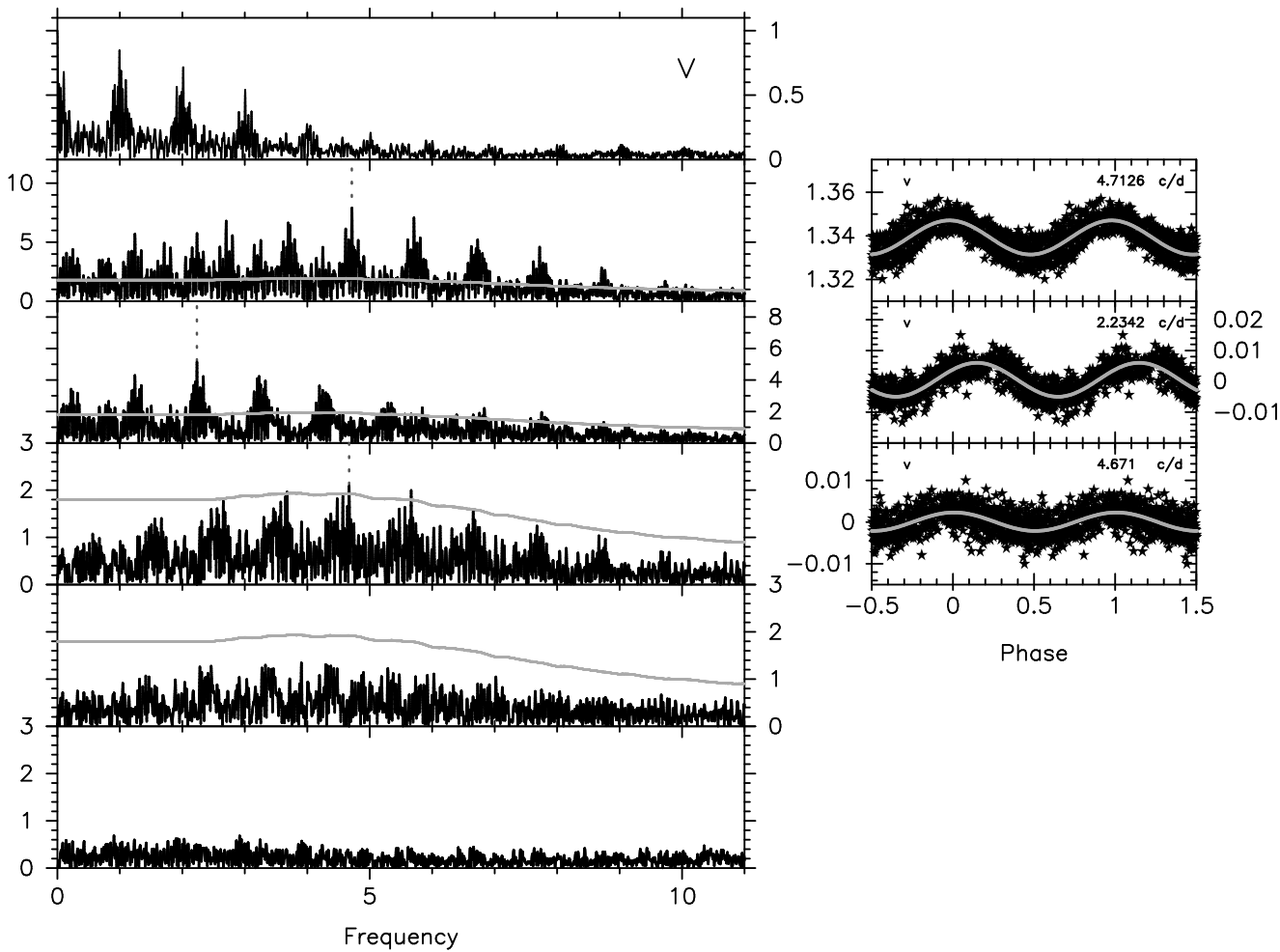}}} 
   \caption{Left panel: Window function (top) and \SCARGLE
   periodograms of V2104\,Cyg at different stages of the frequency
   search in the Str\"omgren {\sl v} filter. The amplitudes are expressed in
   mmag. The detected frequencies \fu$=4.7126\pm0.0004$~\cds,
   \fd$=2.2342\pm0.0005$~\cds and \ft$=4.671\pm0.001$~\cds are indicated
   by dashed vertical lines. The last but one bottom periodogram shows
   the final residuals after prewhitening with \fu, \fd\, and \ft. The
   light grey horizontal lines represent the (S/N$>$3.6)-level. In the
   bottom periodogram we present the variability of the comparison
   stars, illustrating that they are constant at the level of a few
   0.1\,mmag. Right panel: Phase diagrams of the 3 detected
   frequencies in the {\sl v} filter.  Black stars correspond to
   observations and the grey full lines represent the least-squares
   fit. The amplitudes are expressed in mag.}
   \label{scaperiodogram}
   \end{figure*}

\vskip 4mm We first double-checked the reliability of the
\emph{comparison stars}.  The frequency analysis of the combined
timeseries between the two comparison stars did not reveal any
significant term. White noise is around 0.3\,mmag in the {\sl vby}
data and amplitudes of the highest peaks are in the 0.6--0.8\,mmag
interval. The {\sl u} data show a more noisy behaviour. The highest
peaks occur at different frequencies in the different filters,
suggesting a random enhancement. As illustration, we show the
frequency spectrum of the {\sl v} data in the bottom panel of
Fig.~\ref{scaperiodogram}.  Thus, we can state that the two comparison
stars are constant at the level of a few 0.1\,mmag.

\vskip 4mm The frequency analysis of the \emph{V2104 Cyg timeseries}
turned out to be more complex. First, we analysed the SNO and SPMO
datasets separately.
We searched for frequencies in the interval 0.01--25.0~\cds with a
frequency step of 0.01~\cd. We detected similar structures in the
least-squares power spectra and the \SCARGLE periodograms of the
datasets in the different filters and found evidence for two
frequencies centered at \fu=4.71~\cds and \fd=2.23~\cd, or their one-day
aliases. Additional frequency peaks were visible but were below the significance threshold (see below).

Subsequently, a preliminary solution for each dataset was obtained by
means of these two frequencies only and the two datasets were
re-aligned at the same mean brightness level.  In particular, we
obtained the same mean magnitude for the {\sl y} data, within error bars,
while small misalignements were measured in the other data.

In a second step we analysed the aligned full dataset, consisting of 679
datapoints and spanning 42 days. Given the improved frequency resolution
we adopted a frequency step of 0.001~\cd. The spectral window of our timeseries is 
shown in the upper panel of Fig.~\ref{scaperiodogram}.
Besides \fu=4.713~\cds and \fd=2.234~\cd, with amplitudes of
respectively 7.6 and 5.7\,mmag in the {\sl v} filter, we also found in
all filters and using both frequency methods a third term
\ft=4.670~\cd, which has an amplitude of 2.4\,mmag in the {\sl v} time
series.  The amplitudes of all 3 frequencies satisfy the S/N$>$3.6
significance criterion and also the more severe S/N$>$4 criterion
(Breger et al. 1993; Kuschnig et al. 1997; De Cat \& Cuypers 2003) in
the {\sl vby} filters. However, the low-amplitude of \ft\, does not
pass S/N$>$4 in the noisy {\sl u} filter (see last row
Table~\ref{harmonicfit}). The S/N-level was computed as the average
amplitude over a frequency interval with a width of 5~\cds in an
oversampled \SCARGLE periodogram obtained after final prewhitening. We
note that the S/N-level is a factor 1.5 higher than the white noise
detected in the comparison stars (bottom panel
Fig.~\ref{scaperiodogram}).  Comparing the white noise with the
residual time-series of V2104\,Cyg (last but one bottom panel
Fig.~\ref{scaperiodogram}), there might be even more frequencies
present in the signal of V2104\,Cyg, with amplitudes currently below
the significance threshold. For the moment we accept a model with \fu,
\fd\, and \ft. We did not find a direct relation between the three
frequencies, which can be considered independent of each other. To
optimise this triperiodic model we used a code, kindly made available
by Dr.~Jan Cuypers (Royal Observatory of Belgium), that searches
simultaneously a set of frequencies around the input values to find
the multiperiodic combination that fits the data best.  The optimised
solution is: \fu$ = 4.7126 \pm 0.0004$~\cd, \fd$ = 2.2342 \pm
0.0005$~\cds and \ft$ = 4.671 \pm 0.001$~\cd. A similar fit using a
different code (\MTRAP, Carpino et al. 1987) yielded coincident
results. The given accuracy of the frequencies is the average of the
estimated accuracies in the {\sl uvby} filters. The frequency accuracy
in a particular filter $x$ is calculated as $\sigma_{\nu,x} = \sqrt(6)
\sigma_{{\rm std},x} / \pi \sqrt(N) A_{\nu,x} \Delta T$ (Montgomery \&
O'Donoghue 1999), with $\sigma_{{\rm std},x}$ the standard deviation
of the final residuals, $A_{\nu,x}$ the amplitude of the frequency
$\nu$ in filter $x$  and $\Delta T$ the total timespan of the
observations. The solution of the least-squares fit of this
triperiodic model is given in Table~\ref{harmonicfit}. The phases
$\phi_i$ are calculated according to the formula $A_0 + \sum_{i=1,3}
A_i(\cos{2\pi(t-T_0)\nu_i}+ \phi_i)$. Note that the phase values for
\ft\, in the four passbands are in excellent agreement, supporting the
reality of this small amplitude term. The triperiodic fit applied to
the individual datasets of SNO and SPMO yielded the same mean
brightness levels in all filters, i.e.\, the alignment procedure was
insensitive to \ft. The \SCARGLE periodogram of the v filter at
different stages of the frequency search and the phase diagrams with
the 3 frequencies are given in Fig.~\ref{scaperiodogram}. For reasons
of visibility of the peaks around \ft\, we only show the
(S/N$>$3.6)-level, indicated by the light grey horizontal line.

\begin{table*}
\caption{Amplitudes (in mag) and phases (in rad) of the frequencies
   \fu$ = 4.7126$~\cd, \fd$ = 2.2342$~\cds and \ft$ = 4.671$~\cds calculated
   from the {\sl uvby} time-series of V2104\,Cyg by means of a triperiodic least-squares fit. For each filter also the mean
   magnitude difference is given together with the r.m.s. of the residuals. The last row indicates the calculated (S/N$=$4)-level at \ft$ = 4.671$~\cd. Phase = 0 corresponds to $T_0= 2453647.3034$.}
\label{harmonicfit}
\begin{tabular} {l c r  c  rr c rr c rr  c rr }
\hline
\hline
\noalign{\smallskip}
& & & & \multicolumn{2}{c}{{\sl u} light}& & \multicolumn{2}{c}{{\sl v} light} & &
\multicolumn{2}{c}{{\sl b} light}& & \multicolumn{2}{c}{{\sl y} light}\\
\cline{5-6}\cline{8-9}\cline{11-12}\cline{14-15}
\multicolumn{1}{c}{Term} & & \multicolumn{1}{c}{Freq.} & &\multicolumn{1}{c}{Ampl.} &
\multicolumn{1}{c}{Phase} & &\multicolumn{1}{c}{Ampl.} & \multicolumn{1}{c}{Phase} & &\multicolumn{1}{c}{Ampl.} &
\multicolumn{1}{c}{Phase}& & \multicolumn{1}{c}{Ampl.} & \multicolumn{1}{c}{Phase} \\
\multicolumn{1}{c}{} & & \multicolumn{1}{c}{[\cd]} & & \multicolumn{1}{c}{[mag]} & \multicolumn{1}{c}{[rad]} & &
\multicolumn{1}{c}{[mag]} & \multicolumn{1}{c}{[rad]}& &
\multicolumn{1}{c}{[mag]} & \multicolumn{1}{c}{[rad]}& & \multicolumn{1}{c}{[mag]} & \multicolumn{1}{c}{[rad]} \\
\noalign{\smallskip}
\hline
\noalign{\smallskip}
$\nu_1$ & & 4.7126     & &    0.0086 &     0.17 & & 0.0073 &  0.26& & 0.0071  &  0.25 & &  0.0071 & 0.27\\
      & & $\pm$.0004 & & $\pm$.0004 & $\pm$.09 & & $\pm$.0002&$\pm$.04  & &$\pm$.0002&$\pm$.04 &  &  $\pm$.0002&$\pm$.05\\

$\nu_2$ & & 2.2342     & &    0.0098 &     5.41 & & 0.0059 &  5.31& & 0.0058  &  5.28 & &  0.0051 & 5.28\\
      & & $\pm$.0005 & & $\pm$.0005 & $\pm$.05 & & $\pm$.0002&$\pm$.03 & &
      $\pm$.0002&$\pm$.04 &  &  $\pm$.0002&$\pm$.04\\

$\nu_3$ & & 4.671     & &    0.0026 &     6.2 & & 0.0024 &  6.1& & 0.0024  &  6.1 & &  0.0024 & 6.2\\
      & & $\pm$.001 & & $\pm$.0004 & $\pm$.3 & & $\pm$.0002&$\pm$.1 & & $\pm$.0002&$\pm$.1 &  &  $\pm$.0002&$\pm$.2\\
\noalign{\smallskip}
\multicolumn{3}{c}{Mean $\Delta$m ($A_0$)} &&\multicolumn{2}{c}{0.7054$\pm$.0003} &
& \multicolumn{2}{c}{1.3398$\pm$.0001}&& \multicolumn{2}{c}{1.4610$\pm$.0001} &
& \multicolumn{2}{c}{1.5124$\pm$.0001}\\
\multicolumn{3}{c}{Residual r.m.s. [mag]} &&\multicolumn{2}{c}{0.0065} & & \multicolumn{2}{c}{0.0025}&&
\multicolumn{2}{c}{0.0026} & & \multicolumn{2}{c}{0.0028}\\\noalign{\smallskip}
\multicolumn{3}{c}{S/N$=$4 @\ft} &&\multicolumn{2}{c}{0.0038} &
& \multicolumn{2}{c}{0.0022}&& \multicolumn{2}{c}{0.0022} &
& \multicolumn{2}{c}{0.0019}\\
\noalign{\smallskip}
\hline
\end{tabular}
\end{table*}

We also analysed the {\sc hipparcos} data in search for traces of intrinsic
variability. The resulting \SCARGLE periodogram resembled very much the window
function and no frequencies could be detected. The poor spectral window and the less good accuracy of the {\sc hipparcos} data did not allow to detect any of our three small amplitude terms.

\section{{\sl Elodie} and {\sl FIES} spectra: evidence for a Be star \label{spectra}}

\begin{figure}
\centering
\resizebox{0.98\linewidth}{!}{\rotatebox{-90}{\includegraphics{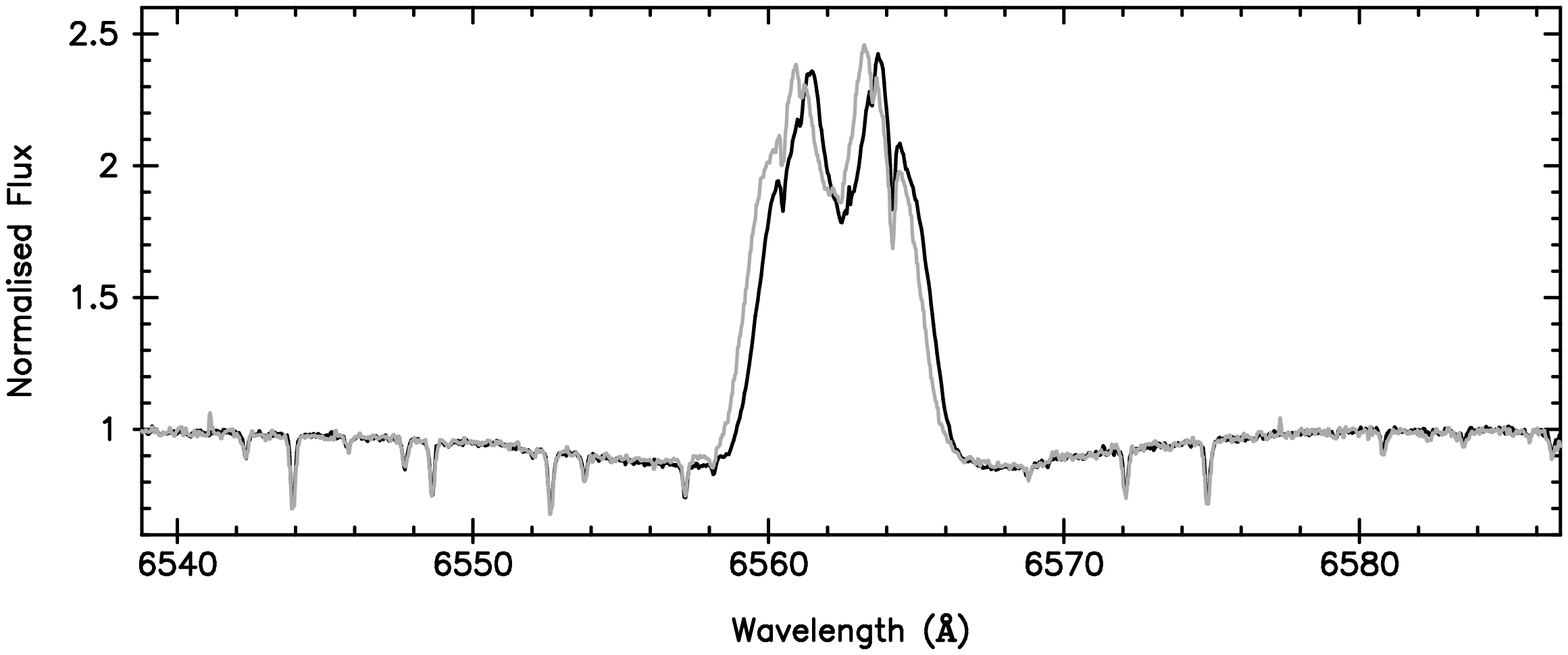}}} \\
\resizebox{0.98\linewidth}{!}{\rotatebox{-90}{\includegraphics{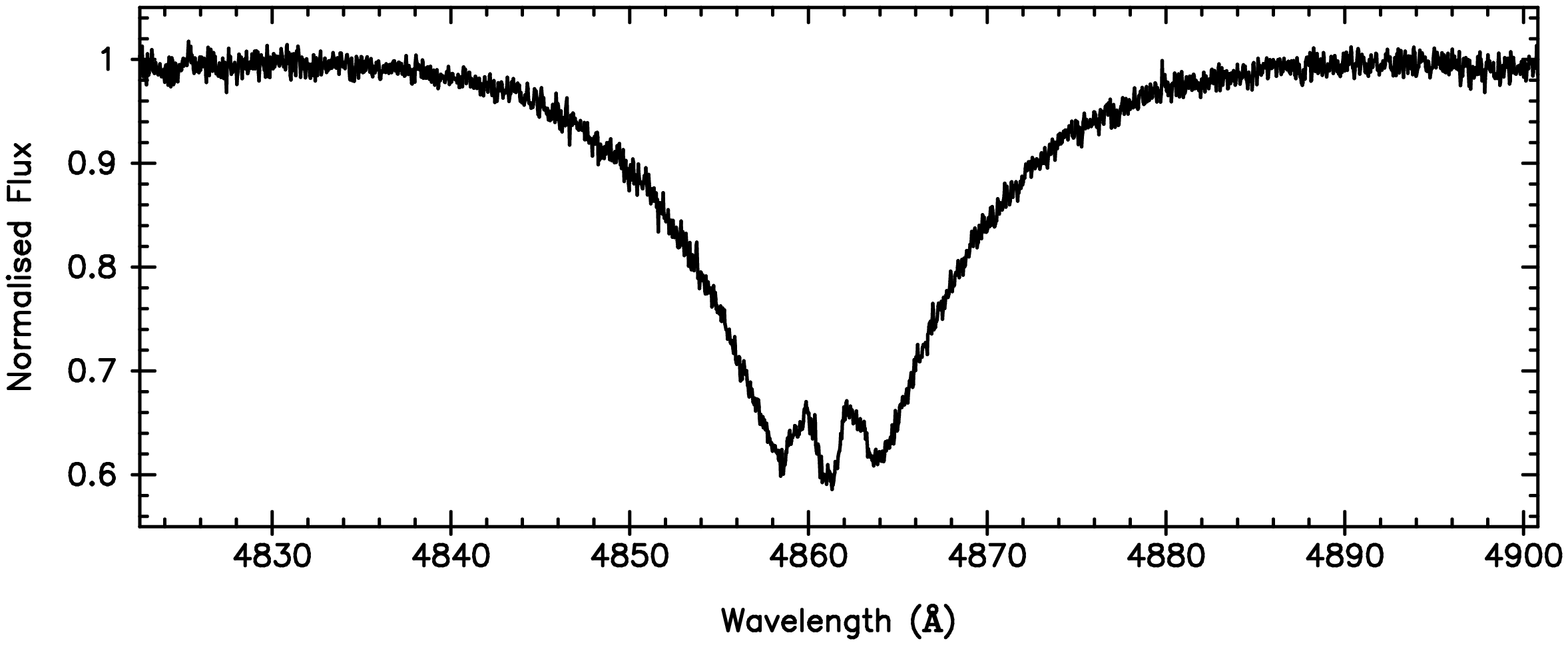}}}
\caption{The \Halpha\, and \Hbeta\, profiles of V2104\,Cyg clearly
  show emission. The black profiles are {\sl FIES} profiles while the
  grey \Halpha\, profile is taken with the {\sl Elodie}
  spectrograph. Between June 2006 and November 2006 the emission peak
  shifted and changed slightly in shape, indications for the dynamical
  character of the circumstellar disk.}
\label{emission}
\end{figure}

In addition to the photometric time-series we obtained 3 spectra 
during 3 consecutive nights in June 2006 with the {\sl Elodie}
spectrograph (R$\sim$45\,000) at the 193 cm telescope of the
Observatoire de Haute Provence (OHP), France, and one spectrum in
November 2006 with the {\sl FIES} spectrograph in medium resolution
mode (R$\sim$45\,000) at the Nordic Optical Telescope located at the
Observatorio del Roque de los Muchachos, La Palma, Spain. We combined
the 3 {\sl Elodie} spectra, after checking that there are no
clear differences visible, resulting in an average
spectrum with signal-to-noise ratio of 70. The signal-to-noise ratio
of the {\sl FIES} spectrum is about 150.

The \Halpha\, and \Hbeta\, line profiles clearly show a double-peaked
emission line (see Fig.~\ref{emission}). Hence, we identify V2104\,Cyg
as a Be star. The emission effect appears to be much stronger in
\Halpha\, than in \Hbeta. In the {\sl FIES} spectra we also see a hint
for emission in the absorption profile of \Hgamma\, but this needs to
be confirmed. A comparison between the \Halpha\, profile (upper panel
Fig.~\ref{emission}) obtained in June 2006 ({\sl Elodie} spectra,
grey) and in November 2006 ({\sl FIES} spectrum, black) shows a radial
velocity shift and small variations in shape of the emission peaks. A
dedicated spectroscopical campaign is necessary in order to study the
dynamical character of the circumstellar disk.

As the line profiles are strongly rotationally broadened, only the
Balmer lines, \HeI\, lines and the absorption line of \MgII~4481 \AA\,
are prominently present and not lost in the noise of the spectra.  To
derive an estimate of v$\sin i$ we calculated theoretical line
profiles for a rotating star by means of the code described by
Schrijvers \& Telting (1999), whereby we assumed a neglegible
contribution to the line shape from NRP. We used limb-darkening
coefficient $\alpha$=0.444, taken from D\'{\i}az-Cordov\'es et
al. (1995) for the physical parameters given in Table~\ref{phys}.  As
Collins et al. (1991) suggest the \HeI~4027 \AA, \HeI~4471 \AA\, and
\MgII~4481 \AA\, profiles as most suitable candidates for the
determination of v$\sin i$, we focussed on these line profiles only.
From the best fits of the observed profiles we estimated v$\sin i$=290
\kms, with formal errors of 10 \kms.

 We calculated the EW of the \Halpha\, profile and obtained
 EW$\sim-6.0\pm0.5$ \AA. The red peak of the emission is higher than
 the violet peak with peak intensities of R=2.42 and V=2.36. 
 The peak
 separation $\Delta V$ between the red and violet peak is about 105
 \kms. As $\Delta V$ is a measure of the radius of the circumstellar
 disk, namely (2 v$\sin i / \Delta V)^2$=$(R_{\rm disk} / R_{\star})$
 assuming a Keplerian rotation (e.g.\, Zamarov et al. 2001), we
 estimate the radius of the circumstellar disk to be about 34 times
 the stellar radius.

Given the double-peaked nature of the emission line, with small
side-lobes on each side (Fig.~\ref{emission}), we expect to see the
circumstellar disk under an inclination angle $i$ between $45\degr$
and $\sim 20\degr$ (e.g.\, Hanuschik et al. 1996).

We compared the {\sl FIES} spectrum of V2104 Cyg with a set of spectra
from the {\sl noao} spectral library (Valdes et al. 2004) and deduced
a spectral type between B5 and B7. Hence we obtain a hotter star than
previously reported (A0, Perryman et al.\,1997; B8, Grenier et
al.\,1999). The misclassification of a late B type star as an early A
type star is not uncommon among fast rotators.  Being a mid-late Be
star, V2104\,Cyg remains an interesting target given that cooler Be
stars are less frequently observed than early-type Be stars.

\section{Physical parameters \label{parameters}}

\begin{figure}
\caption{Possible position of V2104\,Cyg in the Hertzsprung-Russell diagram. The black square
  indicates the position derived from the observed photometric indices while
  the star indicates the position after correction for the contribution of the
  stellar disk. The full
  black lines and the dashed black lines represent the instability strips for
  the SPB and $\beta$ Cephei modes respectively, with a frequency between 0.25
  and 25~\cds and $\ell \leq 4$ computed for Main Sequence models (De Cat et
  al. 2007). The dark grey dotted lines show the ZAMS and the TAMS.}
\resizebox{0.95\linewidth}{!}{\rotatebox{-90}{\includegraphics{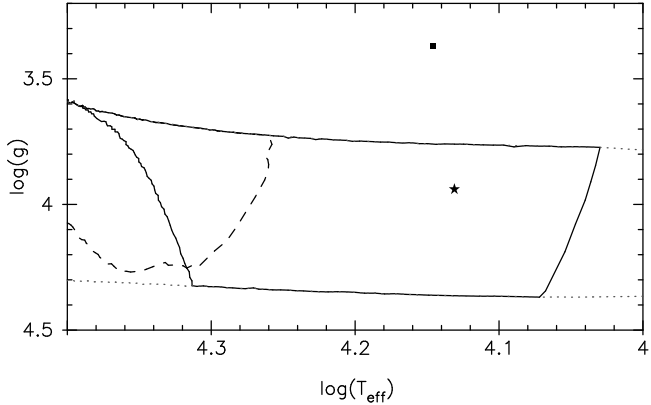}}}
\label{HR}
\end{figure}

First, we calculated $T_{\rm eff}$, $\log g$ and $M_V$ of V2104\,Cyg
from the Str\"omgren indices, whereby using HR~7634 and HR~7692 as
calibration stars.  The observed Str\"omgren indices (all expressed in
mag throughout the text) are $(b-y)=0.000$, $m_1=0.096$ and
$c_1=0.534$. They have been calculated from the mean magnitude
differences reported in Table~\ref{harmonicfit} and from {\sl uvby}
standard photometry of HR~7634. A $\beta=2.681$ value simulaneous to
$uvby$ photometry has been determined from dedicated observations at
SNO.  We obtained dereddened values starting from the above indices
and applying the {\sc TempLogG} method (Kupka \& Bruntt 2001):
$(b-y)_0=-0.057$, $m_0=0.115$, $c_0=0.523$. The interstellar reddening
is non-negligible, $E_{b-y}=+0.057$\,mag, since the star lies only
14\degr\, above the galactic plane.  Additionally, the {\sc TempLogG}
method allowed to derive $M_V=-1.44$\,mag, $\log g=3.37$ and $T_{\rm
eff}=14\,000$~K.  According to these values, V2104\,Cyg is located far
above the instability strip of the SPBs in the Hertzsprung-Russell
diagram (black square in Fig.~\ref{HR}).

As we are dealing with a Be star we expect that the presence of the
circumstellar envelope affects the observed photometric indices. This
might also explain why {\sc simbad} reports scattered values (2.610,
2.648 and 2.705) of the $\beta$ index for our target.  To deduce the
indices from the underlying B star we calculated the transformation
formulae given by Fabregat \& Reglero (1990), which are based on the
assumption that the contribution of the circumstellar disk is
proportional to the EW of the \Halpha\, emission peak. Assuming an EW
of \Halpha\, of $-6.0\pm0.5$ \AA\ we obtained the following new values
of the Str\"omgren indices: $(b-y)_{\star}=-0.01 \pm 0.01$,
$(m_1)_{\star}=0.10\pm0.05$, $(c_1)_{\star}=0.57\pm0.05$ and
$\beta_{\star}=2.73\pm0.01$.  Applying the {\sc TempLogG} method on
these values resulted in dereddened values
$(b-y)_{0,\star}=-0.05\pm0.01$, $(m_0)_{\star}=0.12\pm0.05$ and
$(c_0)_{\star}=0.564$ in addition to $(M_V)_{\star}=-0.49$\,mag, $(\log
g)_{\star}=3.94\pm0.05$ and $(T_{\rm
eff})_{\star}\sim13\,500$~K. After correction for the contribution of
the circumstellar disk, the position of the underlying B star falls
nicely inside the SPB instability strip ($\star$ in Fig.~\ref{HR}).

From the corrected value $M_{V,\star}=-0.49$\,mag we can estimate the
luminosity and the mass of V2104\,Cyg using the relation for Main
Sequence stars between absolute magnitude and luminosity and the
mass-luminosity relation. Assuming that V2104\,Cyg is a MS star we
find: $L/L_{\odot} \sim 127$ and $M/M_{\odot} \sim 4$. Using the
formula $\log (R/R_{\odot})=\frac{1}{2}\log (L/L_{\odot}) - 2 \log
(T_{\rm eff}/T_{{\rm eff}, \odot})$, we estimate $R/R_{\odot} \sim
2.1$. The value of the mass is compatible with the expected values for
a B5--B7 star according to the tabulation of Harmanec (1988), while we
obtained a slightly lower radius (2.1 versus 3.0 $R_{\odot}$).  From
these estimates of mass and radius, the expected upper limit of the
inclination $i<45$\degr (see above), and with v$\sin i$=290$\pm$10 \kms, 
we can derive that V2104\,Cyg rotates at least at 65\% of its
critical velocity, using v$_{\rm{crit}} = \sqrt{G
M_{\star}/R_{\star}}$.  This fact confirms the tight connection
between fast rotation and the Be phenomenon. Moreover, an
interpretation in terms of multiperiodic pulsations would confirm and
strenghten the relation between pulsation and mass loss, and
demonstrate that this relation also exists in mid-late Be stars.
Subsequently, we obtain from the derived v$\sin i$ interval and imposing
v$<$v$_{\rm{crit}}$ a lower limit for the inclination angle:
$i>27$\degr.

Next, from v$\sin i$=290$\pm$10 \kms, a stellar radius in the
interval $[2,3]$ $R_{\odot}$, and an inclination angle between 27\degr
and 45\degr\, we arrive at a (very) rough estimate of the rotational
frequency: $\Omega \in [2.6,6.5]$\,\cd. We notice that the observed
intrinsic frequencies \fu\, and \ft\, lie in this interval. 
It is not uncommon for Be stars that
the rotational period is dominantly present in the light variations
(e.g.\, Balona 1990). We will discuss this possibility further in
Sect.~\ref{discussion}.

We stress that the derived physical parameters tabulated in
Table~\ref{phys} are only a first \emph{estimate} as the contribution
and properties of the circumstellar disk are still unknown and as the
properties derived from the low S/N noise spectra have to be taken
with caution.

\begin{table}
\caption{Estimates of the physical parameters of the Be star V2104\,Cyg, after correction for the contribution of the circumstellar disk. The effective temperature is given in K; the mass and radius are expressed in solar units; v$\sin i$ is given in \kms.}
\label{phys}
\begin{tabular}{cccccc} \hline \hline
$T_{\rm eff}$ & $\log g$ & Mass & Radius & $L/L_{\odot}$ &v$\sin i$ \\ \hline
$13\,500\pm1\,350$ & $3.94\pm0.39$ & $\sim$ 4 & $2.5\pm0.5$ & $\sim 127$ & 290$\pm$10\\ \hline
\end{tabular}
\end{table}

\section{Discussion and conclusions \label{discussion}}
In the light curves of V2104\,Cyg, we detected 3 intrinsic frequencies
with amplitudes below 0.01\,mag: \fu$ = 4.7126$~\cd, \fd$ =
2.2342$~\cds and \ft$ = 4.671$~\cd. We also discovered that this
object is an intrinsically fast rotating Be star.
V2104\,Cyg turned out to be a particularly interesting case as there
are not many precedents for detection of \emph{multiperiodicity} in
\emph{mid-late type Be} stars based on \emph{ground-based photometric}
observations. Indeed, from previous efforts (Cuypers et al. 1989;
Balona et al. 1992) it appears to be not obvious to detect
multi-periodic signals from ground-based photometry in Be stars. Only
recently Guti\'errez-Soto et al. (2006) claimed to have detected
unambiguously several NRP periods in the light curves of NW\,Ser
(spectral type B3) and V1446\,Aql (spectral type B5). If V2104\,Cyg is
a multi-periodic non-radial pulsator as well it would be the third known
 Be pulsator with multiperiodic signals detected in its ground-based light curves.

How can we explain the variability of V2104\,Cyg?  Short periods of
0.2$^d$, like \fu\, and \ft\, are typically detected in early-B type
stars such as $\beta$ Cephei pulsators or early-Be stars, but are
quite unusual in late-type B(e) stars. As V2104\,Cyg lies in the
instability domain of the SPBs, far from the overlap domain with the
$\beta$ Cephei pulsators, we expect to detect longer periods, such as
\fd. 
Rivinius et al. (2003) identified $\ell$=$m$=2 $g$-modes associated
to periods longer than 0.4$^d$ in a selected set of Be stars. Given
that V2104\,Cyg has a spectral type between B5 and B7 it is worth
noting that the NRP stars in Rivinius et al. (2003) have a spectral
type earlier than B6.

 According to Saio et al. (2007) it is likely that \emph{all}
rapidly rotating Be stars show NRP. 
  For fast rotating stars, the comparison of the observed frequencies
$\nu_{\rm obs}$ with the eigenfrequencies of the star $\nu_{\rm
eigen}$ provided by theoretical calculations is not straightforward
(e.g. Berthomieu 1978; Dintrans \& Rieutord 2000). The rotation lifts
the degeneracy of the eigenfrequencies of the modes in $m$ and gives
rise to a perturbation of the frequencies (e.g. Saio 1981; Dziembowski
\& Goode 1992; Soufi et al.\, 1998). These frequency shifts depend on
$m$, the rotation frequency $\Omega$, and other parameters, and can be
very large, especially for $g$-modes.  This perturbation can explain
why the observed frequencies $\nu_1$ and $\nu_3$ of V2104\,Cyg are
larger than expected for stars within the instability strip of SPBs.
Moreover, when the oscillation frequency is of the same order as the
rotational frequency, the whole perturbative approach is invalid and
the familiar structure of modes ($n,\ell$) disappears (Dintrans \&
Rieutord 2000).  Hence, rotation has a significant effect on the
non-adiabatic analysis and mode identification.   Consequently it is not
easy to assign $l$-values to the observed frequencies of a
fast-rotating star, like V2104\,Cyg. In particular, the fast rotation of 
V2104\,Cyg prevents the application of a mode-identification method that does
not account for higher order rotational effects, such as the method of
photometric amplitude ratios made available by Dupret et al. (2003).
 A discussion of the
implications of applying standard mode-identification techniques, used
to identify modes in non-rotating stars, to fast-rotating stars can be
found in Townsend (2003).

It is interesting to point out that for V2104\,Cyg the
amplitudes associated to \ft\, have, unlike the ones associated to
\fu\, and \fd, a similar value in all four filters (see
Table~\ref{harmonicfit}). In general, the amplitudes of NRP modes in B
stars decrease from the {\sl u} filter towards the {\sl vby}
filters. However, similar pulsation amplitudes at different
wavelenghts are predicted for some of the modes of theoretical
models. Several examples can be found in De Cat et al. (2007;
e.g. HD\,14053 and HD\,89688).

An alternative explanation of the observed variability of
V2104\,Cyg might be in terms of a model based on rotational modulation.
In many Be stars the detected period is consistent with the expected
period of rotation of the star (Cuypers et al. 1989; Balona et
al. 1992). 
As seen in
Sect.~\ref{parameters} might \fu\, or \ft\, be related with $\Omega$.
Our dataset did not allow a tight enough constraint on $\Omega$ to
confirm this.  The frequency \fd, however, is much lower than the
expected rotational frequency. Therefore, a model based on rotational
modulation could not explain the light variability of V2104\,Cyg.

The unambiguous
detection of photospheric multiperiodicity in a Be star is in favour of a NRP
interpretation (e.g.\, Porter \& Rivinius 2003). As we detected 3
independent frequencies in the light curves of V2104\,Cyg, the NRP
hypothesis seems currently the most plausible explanation. The
interpretation of pulsational variability in Be stars is complicated
by fast rotation.  Recently, theoretical progress on this topic
has been made by Townsend (2005), who proposed a new explanation for
unstable modes in mid- to late B-type SPB and Be stars in terms of
retrograde mixed modes instead of $g$-modes and by Dziembowski et
al. (2007), who found theoretical backing for the observed frequencies
in the \MOST data of the Be star HD163868 (Walker et al. 2005). The
case study of V2104\,Cyg, a fast-rotating mid-late Be star with 3
excited modes, is another challenge for theoreticians in the game of
matching theory with observations.

A high-resolution spectroscopic time-series of V2104\,Cyg could be a
future, possible observational progress. As LPV are rarely detected
in late-type Be stars (Baade 1989; Rivinius et al. 2003) it would be
an interesting excercise to see if V2104\,Cyg is an exceptional case
with this respect and indeed shows LPV. If present, a line-profile
analysis and subsequent mode-identification will allow to assign
$m$-values to the detected modes. Additionally, a study of the
behaviour of the \Halpha\, emission peak will allow a study of
properties and behaviour of the circumstellar disk.

\begin{acknowledgements}
We thank the referee for providing constructive comments, which
helped to improve our conclusions. We thank Marc-Antoine Dupret and
Josefina Montalban for sharing their opinion on the
mode-identification.  KU acknowledges financial support from a
\emph{European Community Marie Curie Intra-European Fellowship},
contract number MEIF-CT-2006-024476. ER and VC acknowledge financial
support from the Junta de Andaluc\'\i a and the Spanish Direcci\'on
General de Investigaci\'on (DGI) under project AYA2006-06375.
\end{acknowledgements}

\end{document}